\documentclass[useAMS,usenatbib,psfig,times]{mn2e}
\usepackage{psfig}
\usepackage{myaasmacros}
\usepackage{tabularx}
\usepackage{graphicx}
\usepackage[figuresright]{rotating}

\def\etal{et~al.}

\title[A Combined EIS-NVSS Survey Of Radio Sources (CENSORS) II: Infra-red imaging and the $K$--$z$ relation]{A Combined EIS-NVSS Survey Of Radio Sources II: Infrared imaging and the $K$--$z$ relation} 

\author[M.~H.~Brookes \etal]{M.~H.~Brookes$^1$\thanks{Email: Mairi.H.Brookes@jpl.nasa.gov}, P.~N.~Best$^1$, R.~Rengelink$^2$, H. ~J. ~A. ~R\"{o}ttgering$^2$\\ 
$^1$ Institute for Astronomy, Royal Observatory Edinburgh, Blackford Hill,
Edinburgh EH9 3HJ, UK,\\
$^2$ Sterrewacht Leiden, Postbus 9513, 2300 RA Leiden, the Netherlands\\
}
\pagerange{\pageref{firstpage}--\pageref{lastpage}}
\pubyear{2005}
\begin{document}
\label{firstpage}
\def\squig{\sim\!\!}
\def\lesssim{\mathrel{\hbox{\rlap{\hbox{\lower4pt\hbox{$\sim$}}}\hbox{$<$}}}}
\def\gtrsim{\mathrel{\hbox{\rlap{\hbox{\lower4pt\hbox{$\sim$}}}\hbox{$>$}}}}
\def\subsun{\mbox{$_{\normalsize\odot}$}}
\def\deg{\hbox{$^\circ$}}
\def\arcs{\ifmmode {^{\scriptscriptstyle\prime\prime}}
          \else $^{\scriptscriptstyle\prime\prime}$\fi}
\def\arcm{\ifmmode {^{\scriptscriptstyle\prime}}
          \else $^{\scriptscriptstyle\prime}$\fi}
\def\squig{$\sim\!\!$}
\def\Hb{\ifmmode {H\beta}
          \else H$\beta$\fi}
\def\Hg{\ifmmode {H\gamma}
          \else H$\gamma$\fi}
\def\Hd{\ifmmode {H\delta}
          \else H$\delta$\fi}

\def\nii{[N\,{\footnotesize II}]}
\def\oii{[O\,{\footnotesize II}]}
\def\oiii{[O\,{\footnotesize III}]}
\def\Ha{H$\alpha$}
\def\lya{Ly$\alpha$}

\maketitle

\begin{abstract}
\noindent 
The Combined EIS-NVSS Survey Of Radio Sources (CENSORS) is a 1.4 GHz radio survey selected from the NRAO VLA Sky Survey (NVSS) and complete to a flux-density of 7.2mJy. It targets the ESO Imaging Survey (EIS) Patch D, which is a 3 by 2 square degree field  centred on 09 51 36.0 -21 00 00 (J2000).
This paper presents $K$--band imaging of 142 of the 150 CENSORS sources.
The primary motivation for beginning infrared imaging of the sample was to identify the host galaxies of the \squig~~30\% of sources for which the EIS $I$--band imaging failed to produce a likely candidate.
In addition, $K$--band magnitudes allow photometric redshift estimation and  $I - K$ colours aid the identification of host galaxies (which are typically old, red ellipticals).
Of the sources observed in the $I$ and $K$--bands, four remain undetected, possibly indicating high redshifts for the host galaxies, and eight involve complicated radio structures, or several candidate host galaxies, which have yet to be resolved.
Thus, the host galaxy identifications are brought to 92\% completeness.

In conjunction with spectroscopic observations, the $K$--band magnitudes have been aperture corrected and used to establish a $K$--$z$ relation for the CENSORS radio galaxies.
This relation is of interest because of its  variation, at $z > 1$, between radio surveys of different flux-density limit.
Establishing this relation for CENSORS may shed light on the origin of this variation and will allow an appropriate $K$--$z$ redshift estimator for any CENSORS source which remains without a spectroscopic redshift.
It is shown that whilst the $K$--$z$ relation for CENSORS is fainter than that of 3CRR at all redshifts, it agrees well with that of 7C over all redshifts studied.

\end{abstract}


\section{Introduction}

There are many important questions regarding radio-loud AGN astrophysics which remain unanswered.
Just a few examples are:
What causes the radio loudness in some AGN?
How does this vary with host galaxy/black hole mass or environment?
What is the underlying difference between \cite{FR74} Class I and II radio sources (FRI/II)?
How do radio sources evolve individually and as a population and how are these linked to the global star-formation history of the Universe, and structure formation more generally?
One approach to start addressing these questions is the development of large, well-defined samples of radio sources, for which the redshifts are known.

Flux-density limited  surveys offer samples for which the selection effects can be understood, thus making interpretation relatively simple.
However a disadvantage arises from the degeneracy of radio luminosity and redshift due to the flux-density limit.
Because of this it is useful to have several radio surveys, with a range of flux-density limits, which together span a wide range across the radio luminosity--redshift ($P$--$z$) plane.
Producing a radio survey which would extend the current coverage of the $P$--$z$ plane was one of the primary motivations for developing the Combined EIS-NVSS Survey Of Radio Sources (CENSORS; \citeauthor{CENSORS1} 2003; hereafter Paper 1).
This 150 source sample is complete to 7.2mJy at 1.4GHz and covers a field of $2 \times 3$ square degrees  on the southern sky.
The sample was selected from the NVSS \citep{NVSS} in a region overlapping the ESO Imaging Survey Patch D (\citeauthor{EIS}, 1999; EIS).
The EIS $I$--band data ($I \lesssim$ 23.5)  were then used to identify host galaxies for the radio sources.
In this paper  the $K$--band imaging of the CENSORS sample is described.

The first motivation for these observations is to identify host galaxies which were not identified optically.
In Paper 1 around 70\% of the sources were associated with host galaxies which were identified from the $I$--band imaging of the ESO Imaging Survey.
However, in order that this sample may be useful, the remaining host galaxies must be identified.
To detect the remaining host galaxies, it is beneficial to observe at longer wavelengths because  radio source host galaxies are typically hosted by old, elliptical galaxies (e.g. \citeauthor{BLR98} 1998) and these are significantly brighter at longer wavelengths.
This is particularly true at high redshifts where the optical waveband observes the  redshifted rest-frame UV.
In addition, where there was doubt over an $I$--band candidate host galaxy or where there were two galaxies considered likely host candidates, the $K$--band allows a $I - K$ colour to be established and which can be used to aid the identification process.

%K-Z
In addition to observing those sources without host galaxy identifications, there is motivation to observe the entire sample in order to establish $K$--band magnitudes for radio galaxies with the CENSORS sample.
Given the experience of previous radio surveys, it is unlikely that the CENSORS sample can be brought to 100\% spectroscopic completeness in a short period of time and so photometric redshift estimation methods are required in this project.
The $K$--$z$ relation could be used as a first estimate of the redshift for a non-spectroscopically identified host galaxy.
As described above, $K$--band observations of radio galaxies are dominated by emission from the old stellar population. Since this emission does not change as rapidly with evolution of the stellar population as emission at shorter wavelengths, and radio galaxy hosts form a fairly homogenous sample of objects, a tight relation between $K$--band magnitude and redshift exists for radio galaxies (\citeauthor{LL84} 1984, \citeauthor{EalRaw96} 1996, \citeauthor{BLR98} 1998).

However there is  uncertainty in the understanding of the $K$--$z$ relationship's dependence upon radio luminosity.
When the $K$--$z$ relation was first investigated, using solely the 3CRR sample, it was shown that the relation matched the predictions of passive evolution of an old, stellar population and this was interpreted as showing that the high redshift radio galaxies evolved into the low redshift radio galaxies \citep{LL84}.
However subsequent observations have shown that this cannot be the case.
Firstly the high redshift galaxies reside in rich environments, whereas the low redshift galaxies are found to be isolated or in small groups (\citeauthor{Best00} 2000 and references therein).
In addition,  the $K$--$z$ relations for the 3CRR and 6C (the 6CE survey of \cite{6Ce} and the filtered 6C* survey of \cite{6Cstar}) 
samples agree at low redshift but show a mean offset of $\sim 0.6$ magnitudes 
at redshifts of $z > 1$, the 3CRR galaxies being brighter in
$K$ (see \citeauthor{EalRaw96} 1996, \citeauthor{Jarvis01KZ} 2001 and \citeauthor{Inskip02} 2002).
\cite{Willott03KZ} added the 7C survey to this comparison and showed that the 7C and 3CRR surveys are offset in $K$--band magnitude at all redshifts and that small numbers of sources in the low redshift bins of the 6C surveys lead to large errors, thus showing that they too may be consistent with a gradual offset in $K$ as a function of flux-density limit.
The CENSORS survey is approximately 12 times fainter again than the 7C survey (for a typical radio galaxy spectrum of $S \propto \nu^{-0.75}$).
In order to use a $K$--$z$ relation as a first estimate of the redshift for non-spectroscopically identified CENSORS sources, it is therefore necessary to establish a $K$--$z$ relationship for CENSORS, rather than blindly apply applying the relationship developed with other samples of higher flux-density limit.

CENSORS may also be able to help resolve the puzzle.
There are two possible interpretations for the varying $K$--$z$ relation result. 
Firstly, the most luminous radio sources could suffer a significant
contribution, from activity associated with the AGN, to the $K$--band.
Secondly, at high redshifts the more powerful radio sources may be
associated with more massive galaxies \citep{BLR98}. 
CENSORS can be used to test between these since in the first case
contamination could be from a partially obscured nucleus 
or emission line contributions in the 3CR sources \citep{Eales97}.
In this case  little change is expected in $K$ magnitude between the 
6C and CENSORS and this result would have implications for interpretation of
SEDs of 3CRR radio galaxies in terms of stellar populations. 
The second case might be motivated by
suggesting that the most powerful radio sources are powered by the most
massive black holes and, since the black hole mass is roughly proportional
to the galaxy bulge mass \citep{KG01}, hence are hosted by the 
most massive galaxies.
In this case CENSORS galaxies would
be fainter than 6C in the $K$--band and this would have important
implications for the evolution of fueling of RGs (in order that the low and high
redshift results can be reconciled).

The development of the CENSORS sample by identifying host galaxies and targeting them with spectroscopic observations has been carried out over the last four years.
The basic radio data and $I$--band results are presented in Paper 1.
This paper presents the $K$--band imaging and will be referred to as Paper 2, whilst the spectroscopic observations will be presented in \citeauthor{CENSORS3} (in prep.; Paper 3).
Note that  some of the results of the spectroscopic observations are referenced in this paper in order that aperture corrections to the $K$--band magnitudes may also be presented here.
In addition to these three data papers, an investigation of the cosmological evolution  of radio sources has been carried out. This was the primary scientific goal of developing this sample  and in \citeauthor{CENSORS4} (in prep.; Paper 4) a study of  that topic will be presented.

The layout of the current paper is as follows.
Section \ref{imaging} describes the observations, data reduction methods and photometric analysis of the data.
Section \ref{results} presents the results of these observations and includes tabulated photometric measurements and the $K$--band images of a sub-sample of the CENSORS sources.
Section \ref{apcorr} describes the aperture corrections to the observed $K$--band magnitudes for those sources with a spectroscopic redshift (presented in Paper 3). 
Section \ref{kz} presents the CENSORS $K$--$z$ relation for radio galaxies and a summary of results and conclusions is given in Section \ref{summary}.
Throughout this paper the following cosmological parameters are adopted: $H_{0} = 70\rm{kms}^{-1}\rm{Mpc}^{-1}$, $\Omega_{M} = 0.3$ and $\Omega_{\Lambda} = 0.7$.

\section{Imaging}
\label{imaging}
\subsection{Observations}

The $K$-band imaging of 142 of the 150 CENSORS sources was carried out  over several runs using UFTI on UKIRT (2001, 2002), ISAAC on the VLT (2002) and IRIS2 on the AAT (2003, 2004), the latter three runs being done in service mode. 
UFTI has a 1024x1024 HgCdTe array and a plate scale of 0.09\arcs/pixel, giving a field of 92\arcs by 92\arcs; ISAAC has a 1024x1024 Rockwell detector, with a  pixel scale of 0.1484\arcs /pixel and covering a field of 2.5\arcm x 2.5\arcm, IRIS2 again has a 1024x1024 Rockwell HAWAII-1 HgCdTe  detector with 0.446\arcs /pixel, giving a field of view of 7.7\arcm x 7.7\arcm.
All observations were made using a 9-point jitter pattern of offset exposures in order to produce a good flat-field and to allow removal of the rapidly varying infrared sky background. Typical seeing for these observations were: UKIRT 2001: 0.8\arcsec; UKIRT 2002: 0.65\arcsec; VLT 2002: 0.5\arcsec, AAT 2003: 0.8\arcs, AAT 2004: 1\arcs-2.5\arcs.
The $K^{\prime}$ improves on the $K$--band (for non-$IR$ optimized telescopes) by being slightly narrower in order to reduce thermal background from warm ground-based telescopes.
All the telescopes (other than UKIRT)  had a $K^{\prime}$ (or similar) filter available and this was used. 
However since photometric calibration was done using $K$ magnitudes, and since the colour dependence of the correction from $K^{\prime}$ to $K$ ($\le 0.01\rm{mags}$) is much smaller than the errors, the difference is ignored hereafter, and all filters are referred to  as $K$ for simplicity.
The details of these runs are given in Table \ref{Kobs_table} and are associated with a run number, given in this table, by which they will be referred to later.

\begin{scriptsize}
\begin{table}
\begin{center}
\begin{tabular}{|l|l|l|c|} \hline
Run      &Date          		&Telescope      &Nights\\
		&			&(Instrument)			&Photometric\\ \hline
1		&20010203 - 06  	&UKIRT		(UFTI)		&4\\
2*		&20020419 - 26		&UKIRT		(UFTI)		&2\\
3		&20020316,20020415 	&VLT		(ISAAC)		&2\\
4		&20030606,20030613	&AAT		(IRIS2)		&0\\
5		&20030120 - 23		&AAT		(IRIS2)		&0\\
6		&20040401 - 20040403	&AAT		(IRIS2)		&0\\
\hline
\end{tabular}
\caption
[$K$-band imaging observing runs for CENSORS]
{Details of  observing runs for $K$--band imaging of CENSORS. *Included in this data set are two observations made in service mode: CENSORS 51 and 98 on 20021127 and 20030204 respectively\label{Kobs_table} }
\end{center}
\end{table}
\end{scriptsize}

Observations of standard stars were made on photometric nights.
These observations were generally five point jitter patterns and were reduced by the same prescription as the science targets.

\subsection{Data Reduction}
Data reduction was performed identically for observations from all observing runs. It was done using the \verb+DIMSUM+ task\footnote{DIMSUM is the `Deep Infrared Mosaicking Software' package, developed by Eisenhardt, Dickinson, Standford and Ward} in the \verb+IRAF+ data reduction package (\citeauthor{IRAF86} 1986; \citeauthor{IRAF93} 1993). Dark frames were subtracted. A flat-field was produced by averaging over images from the full jitter pattern, and was used to normalize the images. These images were sky-subtracted and cosmic rays were removed. Following this they were registered and combined in the \verb+reduce+ task by reference to the brightest unresolved objects in the frame. A bad pixel map was produced and used to remove bad pixels in each image, before the final combined image was produced.

\subsection{Astrometry and Host Galaxy Identification}
Accurate astrometry is essential in order to correctly identify the counterparts to the CENSORS radio sources. This was done using IDL by comparing bright, but unsaturated stars with those in the EIS $I$--band catalogue. The procedure in IDL fits pixel scale, image rotation and reference RA and Dec.
The astrometry is typically certain to $\lesssim$ 0.1\arcs relative to the $I$--band images. The $I$--band images themselves are matched to the radio frame with an astrometric accuracy of within 0.2\arcs \citep{EIS}. However, note that for those sources for which a only a faint detection has been made the measured position may be much less certain (\squig~~1\arcs).

\subsection{Photometry}

On photometric nights the $K$--band magnitudes of standard stars were used to determine the zero point of the images for the night.
The standards were taken from the Faint UKIRT JHK Standards list \citep{UKIRTFS} and  were chosen on the basis of their brightness and proximity to the CENSORS target field.
They include: FS3, FS19, FS20, FS27, FS103, FS121, FS124, FS125, FS126, FS128.
The calculated  zero point was then used to calculate the magnitude of the target sources from their counts.
The zero point varied by \squig~~0.05 mags between different standard stars and this was taken to be the error in this figure.

The zero point may also determined for non-photometric images using stars within the field of view from which the $K$--band magnitude is known.
These magnitudes are obtained either with reference to previous photometric images or from the Two Micron All-Sky Catalog of Point Sources (2MASS\footnote{This work makes use of data products from the Two Micron All Sky Survey, which is a joint project of the University of Massachusetts and the Infrared Processing and Analysis Center/California Institute of Technology, funded by the National Aeronautics and Space Administration and the National Science Foundation.}).
When 2MASS was used, only stars brighter than K = 14.3 were used for calibration and
the zero point was taken to be the  average of those determined for the calibration sources.
The error was taken to be the standard deviation of those zero points.

\smallskip
\noindent
{\bf Aperture radius}

Since the observed sources span a  wide range in redshifts (c.f. Paper 3), apertures of the same radius on the sky, will not correspond to the same physical radius and so, ultimately, it is desirable to correct the measured $K$--band magnitudes to a standard physical aperture, on the basis of the redshift of the source. 
In addition, more ad hoc problems occur, such as neighbouring sources, which prevent a large aperture measurement being possible.
In order to achieve this in a simple and yet flexible manner, the photometric analysis has been performed for several apertures of different radius, where possible. 

For targets observed at UKIRT, 1, 1.5, 2.5 and 4.5\arcs radius apertures were used.
Targets observed at the AAT were brighter and therefore, typically, closer. 
A combination of the fact that these sources were more likely to be extended than the UKIRT targets and the fact that IRIS2 has larger pixels meant that  a 1\arcs aperture was usually unpractical. 
So these targets were measured with only the 1.5, 2.5 and 4.5\arcs apertures.
Some targets required smaller apertures due to the presence of nearby sources or because they were very weak and this was the only way to obtain an accurate measurement.

\smallskip
\noindent
{\bf Magnitudes and errors}

The counts detected for a source were determined using the \verb+phot+ task in \verb+IRAF+. The \verb+phot+ task measures counts in circular apertures and calculates sky counts from an annular aperture. 
It was applied to the target, then \squig~10 identical apertures are placed on blank patches of sky near the target. 
These data were then used to measure the Gaussian error in the source counts, the Gaussian error in the sky counts and the standard deviation of the residual (post-sky subtraction count) in the blank apertures. 
This residual tells us whether the error in the sky background is dominated by the variance of the sky background within an aperture or by the varying sky background over the image. 
If the residual error is greater or of the order of the sky variance within an aperture, which in almost all cases it was, then it is combined in quadrature with the source counts error (as it already includes the effect of sky variance within an aperture).
If it is negligible, the sky variance within an aperture is used instead.

\section{Results}
\label{results}

Of the 150 CENSORS sources, 102 likely host galaxy detections were claimed in \cite{CENSORS1}. 
However of these two had more than one likely host galaxy candidates; in addition subsequent spectroscopy (see Paper 3) has shown that two candidates were stars and therefore misidentified.
Figure \ref{kband_ps} presents the $K$--band images for those sources for which the $K$--band data have provided a host galaxy identification, those which  remain undetected, and  those for which there are complications  identifying the host. 
Some images have been Gaussian smoothed for clarity.
The results of the photometry are given in Table 2.
For those sources undetected in $K$, a 2--$\sigma$ limit based on a  1\arcs radius aperture is given.
Positions based upon the $K$-band images are given, though it should be noted that for weak detections these may only be accurate to 1\arcs.

\subsection{Using $I$--$K$ as an indicator of host galaxy type}

\begin{figure}
\psfig{file=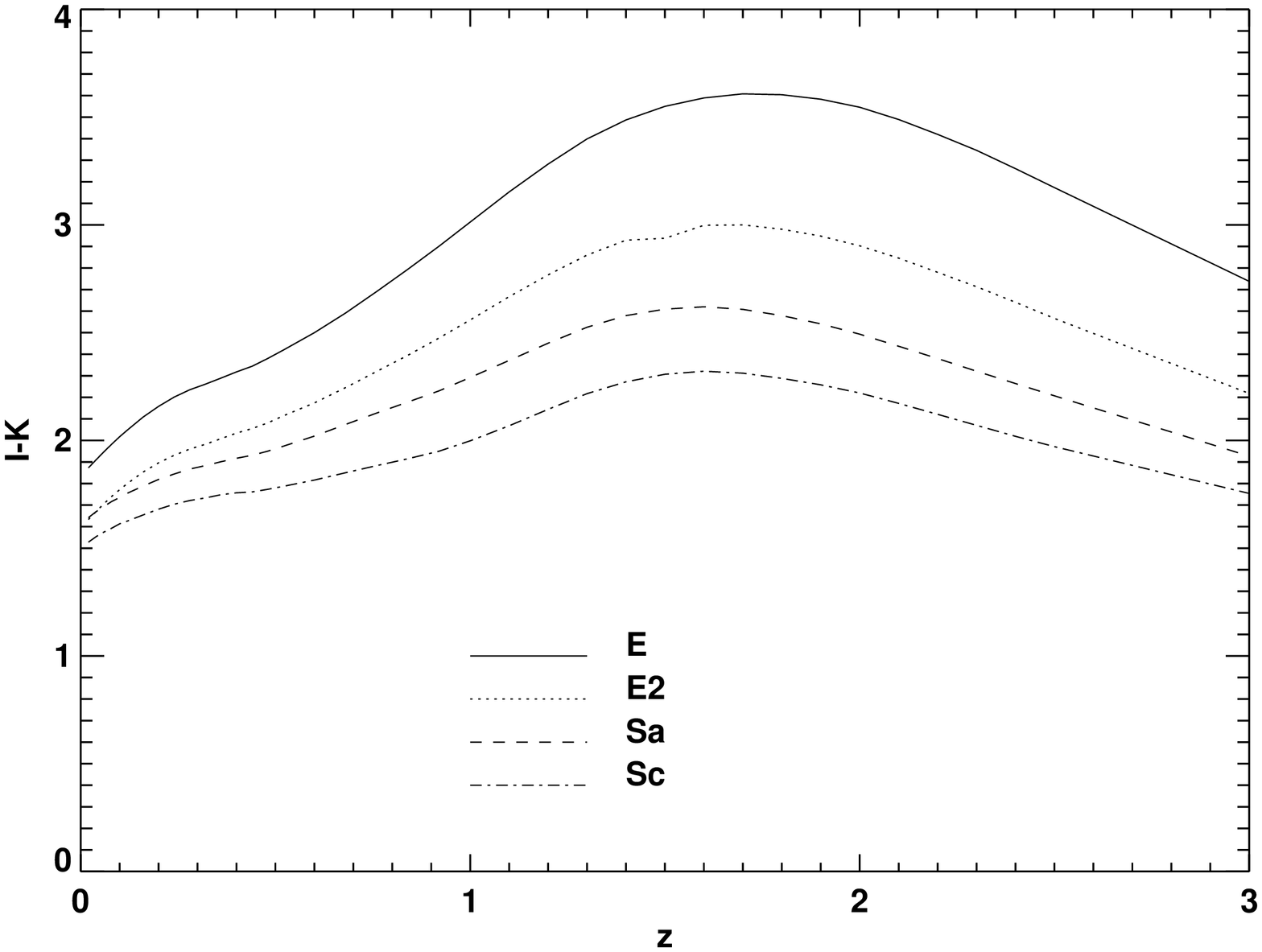,angle=90,width=\linewidth}
\caption
[The variation of $I$--$K$ with redshift]
{
The variation of $I$--$K$ for four galaxy models as derived from the evolutionary and $K$-corrections of \cite{pogg97}.
\label{I-K_pogg}
}
\end{figure}

In the following discussion of individual sources, the  $I$--$K$ colour is used as an indicator that a given galaxy is the correct host galaxy ID.
`Large' values of $I$--$K$ are used as evidence that a galaxy has an old stellar population.
This is based on the fact that radio galaxies are generally hosted by old, giant elliptical galaxies.
At any given redshift these are generally the reddest galaxies because they have the oldest stellar populations.
This is demonstrated in Figure \ref{I-K_pogg} which plots, based upon the $K$ and evolutionary corrections of \cite{pogg97}, the predicted $I$--$K$ for four types of galaxy.
These galaxy types are two elliptical models, E and E2, which differ in their e-folding times for the star-formation rate (1Gyr and 1.4Gyrs respectively) and two spiral models (see \citeauthor{pogg97} 1997 for details).
At a given redshift the elliptical galaxies always have redder $I$--$K$ colours, however the actual value of $I$--$K$ which is considered red varies with the redshift of the galaxy (or the $K$-band magnitude via the $K$--$z$ relation).
For example, at $z = 0.5$, $I$--$K$ = 2.5 would correspond to a red elliptical, but at $z = 1.5$ such a colour would be typical of a spiral galaxy.
Given $I$--$K$ and an estimate of the redshift this plot can be used to check for consistency with a giant elliptical galaxy, thus giving further evidence that the candidate is, indeed, the host.
Note that even redder colours are possible if the star formation timescale of the ellipticals is shorter.
The $I$--$K$ colours ought to be calculated via $I$ and $K$--band magnitude measured with equal aperture radii.
As the EIS $I$--band photometry was derived using SExtractor \citep{SEX}, and the $K$--band magnitude used relatively small aperture radii, the colours obtained give a general, rather than highly accurate, indication of the nature of the host galaxy. 

\subsection{Notes on particular objects}

The following are notes regarding particular targets (they include some references to Paper 3).

\begin{table*}
\centerline{\psfig{file=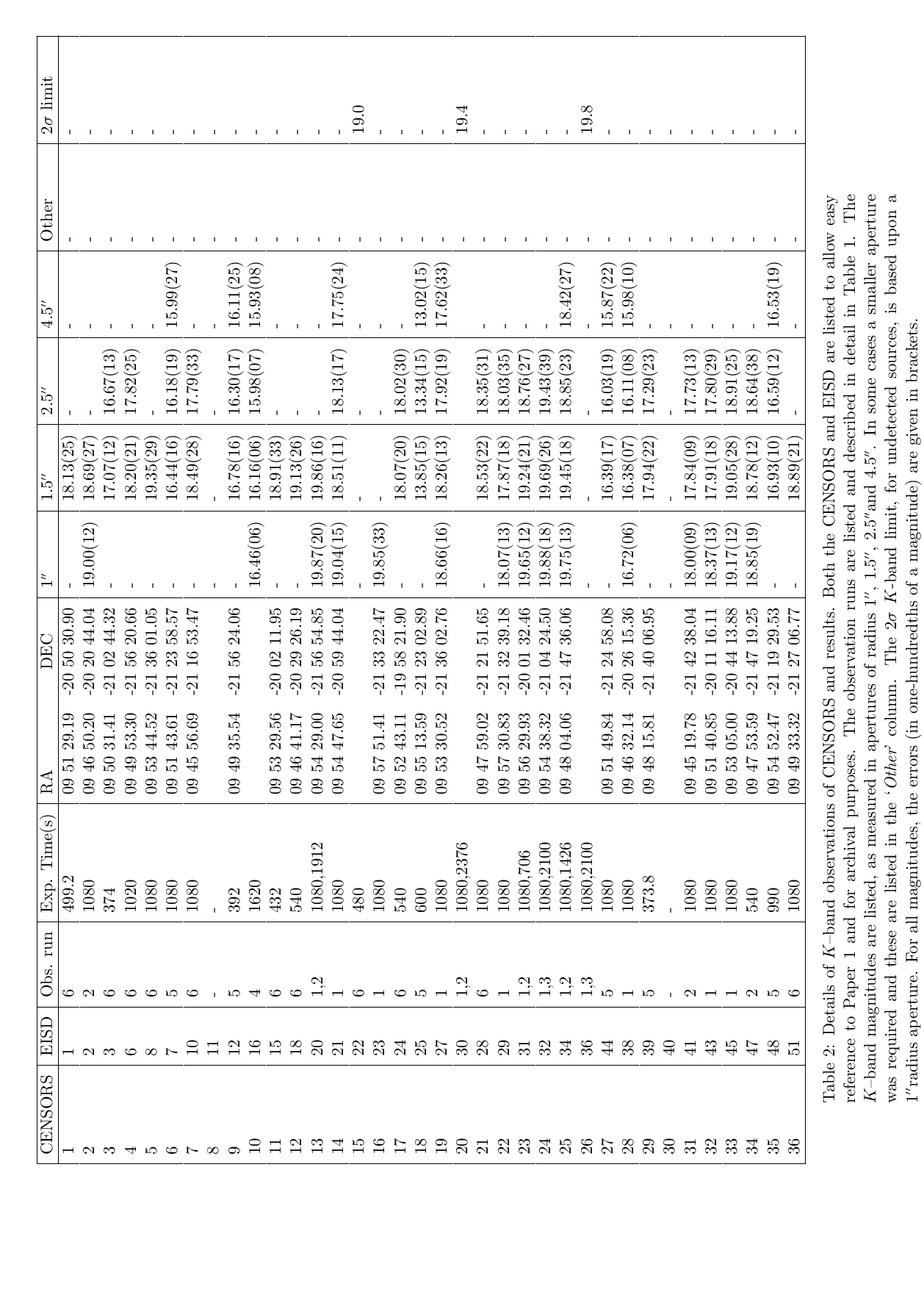,angle=0,width=\linewidth}}
%\caption{\label{phot_table}}
\end{table*}

\begin{table*}
\centerline{\psfig{file=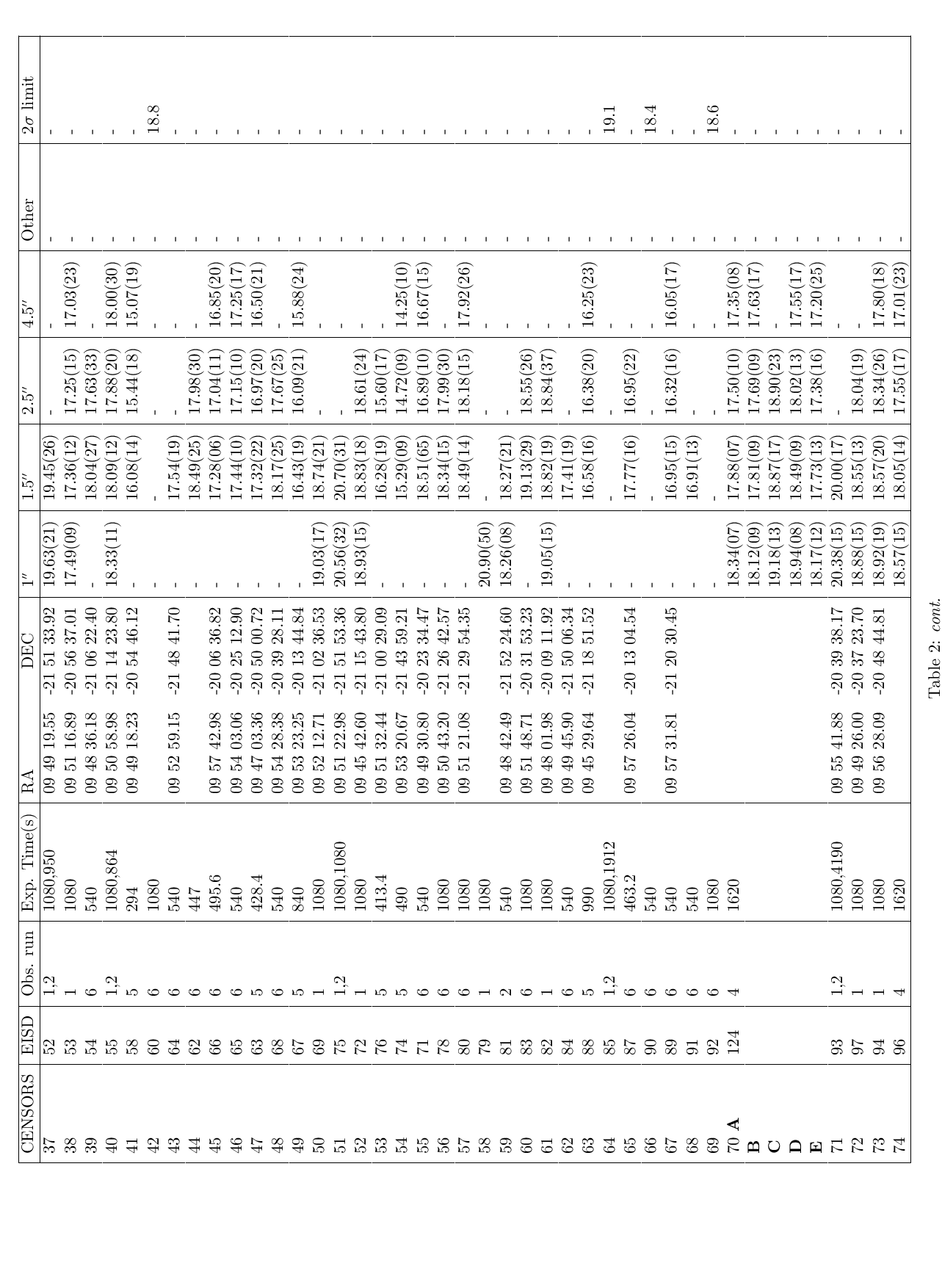,angle=0,width=\linewidth}}
\end{table*}

\begin{table*}
\centerline{\psfig{file=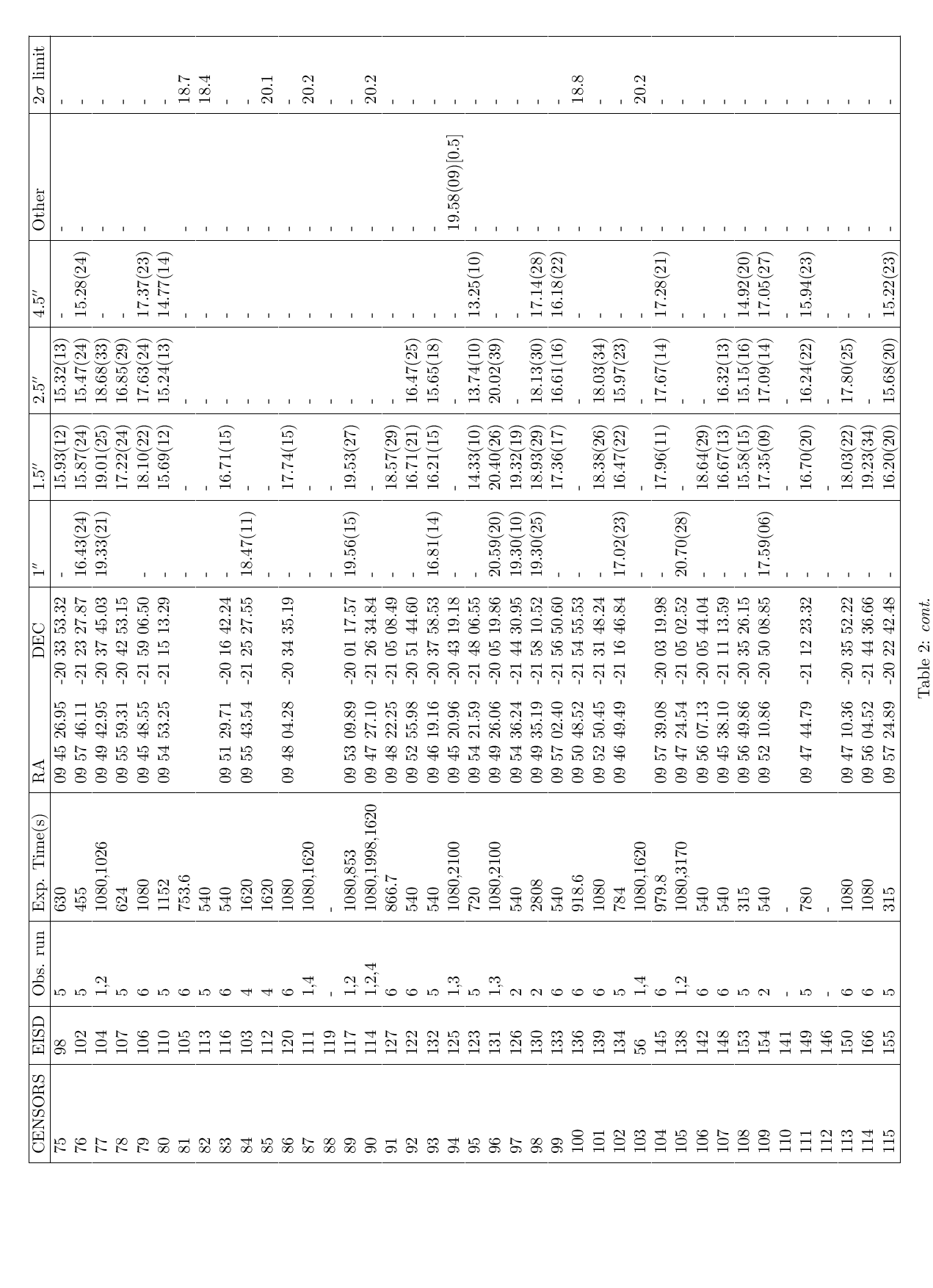,angle=0,width=\linewidth}}
\end{table*}

\begin{table*}
\centerline{\psfig{file=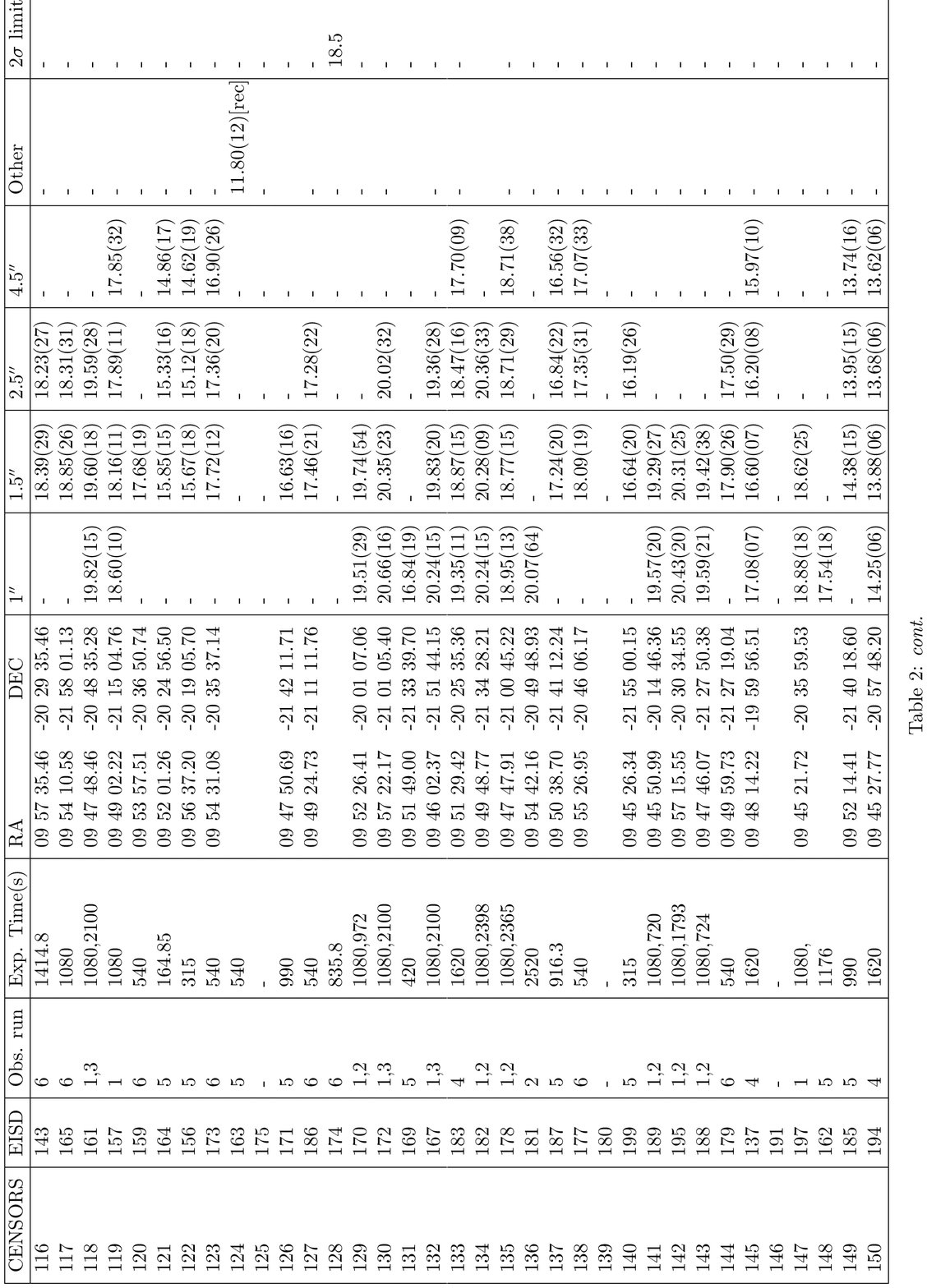,angle=0,width=\linewidth}}
\end{table*}

\setcounter{table}{2}

\noindent
{\bf CENSORS 2:} The faint $I$--band ID is confirmed.

\noindent
{\bf CENSORS 4:} The northern of the two $I$--band candidates has been spectroscopically confirmed as the host galaxy.

\noindent
{\bf CENSORS 7:} This faint ID is confirmed in the $K$--band and spectroscopically.

\noindent
{\bf CENSORS 13:} An ID for the host galaxy has been found on the SE component. As discussed in Paper 1, it is likely that the NW radio components are not associated with this and that we have  a single source. In that case those two NW sources do not have sufficient flux to be included in the sample.
The radio flux density given for this source, in Paper 1,  is for the SE component only.

\noindent
{\bf CENSORS 14:} The faint $I$--band ID is confirmed.

\noindent
{\bf CENSORS 20:} There is a potential $K$--band ID which is a red source associated with the northern radio component. However spectroscopy (see Paper 3) reveals that emission lines are associated with the centre of the southern component, indicating that the true host is located there. The appropriate magnitude here is then the limiting magnitude at that point, which is 19.4.

\noindent
{\bf CENSORS 28:} $I$-$K$ \squig~~3. For a low--$z$ galaxy, which this must be from its $K$--band magnitude, this is a very red colour which reinforces this candidate as the likely host galaxy.

\noindent
{\bf CENSORS 31:} The western optical source is a star. The eastern candidate has not been spectroscopically confirmed but has 
$I$-$K$ \squig~~2.7 which is not inconsistent with an AGN host at its $K$--$z$ estimated redshift.

\noindent
{\bf CENSORS 32:} The faint $I$--band ID is confirmed.

\noindent
{\bf CENSORS 37:} The faint $I$--band ID is confirmed both in $K$ and spectroscopically.

\noindent
{\bf CENSORS 38:} Whilst this candidate was not found to be likely according to the analysis of \cite{CENSORS1}, subsequent spectroscopy shows that it is a quasar and is therefore the correct identification for the radio source host galaxy. 

\noindent
{\bf CENSORS 42:} The faint $I$--band ID is not detected in this (shallow) $K$--band observation but has a single emission line (see Paper 3) and is therefore correctly identified as the AGN host.

\noindent
{\bf CENSORS 43:} The faint $I$--band ID is confirmed both in $K$ and spectroscopically.

\noindent
{\bf CENSORS 47:} The $I$--band ID is confirmed both in $K$ and spectroscopically.

\noindent
{\bf CENSORS 57:} Whilst no spectroscopic confirmation has been obtained, the red, $I$-$K$ \squig ~~3.7 colour, of this galaxy provides additional evidence that it is indeed the host galaxy.

\noindent
{\bf CENSORS 58:} There is a very tentative detection of a host galaxy for this source, however, a nearby source dominates emission.

\noindent
{\bf CENSORS 60:} The $I$--band ID is confirmed in $K$ and has $I$-$K$ \squig~~4.5. This red colour is indicative of a $1 \le z \lesssim 2$ elliptical, consistent with an AGN host.

\noindent
{\bf CENSORS 62:} The faint $I$--band ID is confirmed in $K$ and has $I$-$K$ \squig~~3, which is red for its expected redshift.

\noindent
{\bf CENSORS 67:} The $I$--band ID is confirmed in $K$ and has $I$-$K$ \squig~~2.3. 

\noindent
{\bf CENSORS 69:} There is a faint source on the $I$--band image. However, it was not identified as a likely candidate. There is no detection in the $K$--band so a $K$--band limit is measured.

\noindent
{\bf CENSORS 70:} This is a large extended radio source for which is was not possible to easily associate a host galaxy based on the I band imaging from EIS.
K band imaging allows candidates to be selected on the basis of colour. These are presented in Table \ref{eis124_phot}.
Since CENSORS 70 is a fairly large radio source (3$\arcm$) it is likely that it is at relatively low redshifts (at $z > 1$ this angular size corresponds to \squig~~1Mpc).
Thus candidates B and E, which  have $I - K$ typical of elliptical galaxies at $z < 1$ are, by this argument, the most likely candidates.
This issue is unlikely to be resolved without spectroscopic follow up of all these candidates.

\begin{scriptsize}
\begin{table}
\begin{center}
\begin{tabular}{ll}\hline
Candidate	&\squig~~I-K \\
		& \\\hline
A		&3.25\\
B		&2.33\\
C		&$>$4.1\\
D		&4.07\\
E		&2.14\\
\hline
\end{tabular}
\caption{Photometry for CENSORS 70. I-K colours are approximate (since no aperture correction has been made) and are based on the 1.5\arcs aperture\label{eis124_phot}}
\end{center}
\end{table}
\end{scriptsize}

\noindent
{\bf CENSORS 74:} There were no likely candidates according to the analysis of \cite{CENSORS1}, however there are two optical sources visible in the $I$-band image which might be associated with radio source. $K$-band imaging reveals that the more eastern of these two is a mildly red ($I$-$K$ \squig~~2.9) source and is adopted as the candidate.

\noindent
{\bf CENSORS 81:} The host galaxy is detected in $K$ and has been spectroscopically confirmed.

\noindent
{\bf CENSORS 82:} The identification came from the $I$-band, but the spectrum shows this to be a star. There is no host galaxy candidate from the $K$-band.

\noindent
{\bf CENSORS 84:} As discussed in \cite{CENSORS1} this radio source (EISD 103) was associated with EISD 73 and EISD 151. It is believed that EISD 73 and 151 were correlated noise in the NVSS data thus leaving EIS103. 
The higher resolution maps then revealed this to be two radio sources leaving the possibility that either they were associated, and would therefore have sufficient combined flux to be included in CENSORS, or that they were unassociated.
The $K$--band imaging reveals that there is no red galaxy between the two radio sources that might indicate that they are associated and are radio lobes. 
There is a candidate host galaxy associated with the western radio source. This is taken to be the correct host galaxy for this source and it continues to be included in the CENSORS sample on the assumption that the eastern source is associated with it.

\noindent
{\bf CENSORS 86:} The host galaxy is detected in $K$ and,  whilst not spectroscopically confirmed, $I$-$K$ = 5.3 and this is typical of the old elliptical galaxies which host AGN.

\noindent
{\bf CENSORS 89:} There are two K band candidates for this source. The eastern identification emits an [OII] line and shows strong H and K absorption features (see Paper 3).

\noindent
{\bf CENSORS 90:} As discussed in \cite{CENSORS1} CENSORS 90 and 103 (EISD 114 and 56 respectively) may be associated (CENSORS 10/EISD 16 is nearby but is clearly associated with  a double radio source).
Figure \ref{censors103_fig} shows the $I$ and $K$-band images with the radio contours. 
Component A is CENSORS 10. 
It was suggested that the optical counterpart to component F was evidence that it was a single source, too weak to be included in the CENSORS sample. 
This is consistent with the counterpart being a red source, detected in the $K$-band.
B may correspond to CENSORS 90 while component E corresponds to CENSORS 103; component D has a counterpart and causes the NVSS map for CENSORS 103 to be extended to the NE. 
However, neither component B or E has a counterpart in either the optical or near-IR, whereas component C does and lies between the two, raising the possibility that B, C and E are associated.

\noindent
{\bf CENSORS 94:} The off-axis NE source is believed to be the host galaxy. The other candidate is a star and the NE source has $I$-$K > 4$.

\noindent
{\bf CENSORS 97:} $I$-$K$ \squig~~ 3.2 confirms this as the likely candidate.

\noindent
{\bf CENSORS 103:} See CENSORS 90.

\noindent
{\bf CENSORS 104:} There is no spectroscopic confirmation but this has $I$-$K$ \squig~~5.3.

\noindent
{\bf CENSORS 105:} This red (I-K $\gtrsim$ 3.3) host galaxy candidate is confirmed by extended \lya~~emission (Paper 3).

\noindent
{\bf CENSORS 107:} This galaxy has $I$-$K$ $\gtrsim$ 2.6, and has been shown spectroscopically to be at $z$ \squig~~0.5 (Paper 3). The spectrum is  of an old galaxy, and the colour is red for that redshift, suggesting that this is indeed the host.

\noindent
{\bf CENSORS 109:} With $I$-$K$ $\gtrsim$ 2.9, this is a relatively red galaxy for its expected redshift. 

\noindent
{\bf CENSORS 115:} This galaxy has $I$-$K$  $\gtrsim$ 2.6. Again the spectrum shows a well defined 4000\AA~ break and G-band absorption. This indicates that the galaxy is old and therefore a reasonable identification for the host.

\noindent
{\bf CENSORS 126:} There were two $I$-band identifications for this object one of which has been identified as a star via spectroscopy. The second candidate is taken as the host galaxy, however the radio emission does not clearly associate with this optical/near-IR source by eye. The second candidate is apparently unresolved in the $K$--band image, however it is very close to the star making it difficult to tell.
It is possible that the true ID could be hidden behind the star.

\noindent
{\bf CENSORS 133:} The red colour, $I$-$K \gtrsim 4$, of this galaxy, in addition to its position, coincident with the radio source, makes it the likely host.

\noindent
{\bf CENSORS 135:} The candidate found on one component is confirmed in the $K$--band. It has a colour of $I$-$K$ $\gtrsim$ 2.5, which is potentially blue.

\noindent
{\bf CENSORS 136:} A red source appears in the $K$-band. It is somewhat offset from the main radio component, but may be reasonable if the  radio emission to the south of the main component is associated with this source.

\noindent
{\bf CENSORS 148:} This host galaxy is very close to a brighter nearby object. This means that the quoted error on the magnitude may be optimistic.

\noindent
{\bf CENSORS 150:} There is no clear candidate associated with the source. For now the brightest galaxy is taken to be the host galaxy.

\subsection{Summary}

142 of the 150 CENSORS sources have been targeted in the $K$--band.
There are eighteen sources for which a host galaxy is  not clearly detected.
However, of these, six are detected in the $I$--band. The remaining twelve are CENSORS 20, 26, 58, 64, 69, 70, 82, 85, 87, 90, 103 and 150.
Some of these sources have complications, for example, as discussed in Paper 1, CENSORS 82 may be associated with CENSORS 66 (which has an $I$--band identification) and CENSORS 90 may be associated with CENSORS 103.
Whilst CENSORS 20 has no identified host galaxy,  a spectrum taken  at the radio source position shows  two emission lines ($z$ = 1.38).
CENSORS 85 has no clear host galaxy identification but this may be due to the complexity of the radio map rather than lack of depth in the optical and infrared images.
CENSORS 58 is very close to  a bright star, making it very difficult to detect the host; CENSORS 150 has no obvious candidate host and  CENSORS 70 has several candidates (though spectroscopic targeting ought to resolve the correct host).
Low frequency radio data, which has recently been obtained at 610 and 325MHz, may help to resolve issues where interpretation of the radio data is difficult.
This leaves four sources which are not detected  and have no other complications:  these are likely to be cases of high redshift host galaxies, and  deeper infrared images are required.

There may also be, as yet unidentified, problems due to lensing of radio sources. It may be possible that a radio source has been associated with a relatively low redshift galaxy (and subsequently its redshift) when in fact the low redshift galaxy is lensing the radio source which is actually at a high redshift. This could have consequences when the redshifts ascribed to radio sources are used in the modelling of the radio luminosity function.

\section{Aperture Corrections}
\label{apcorr}

\begin{scriptsize}
\begin{table}
\begin{center}
\begin{tabular}{|l|l|l|l|l|l|l|l|}
\hline
CEN   & EIS    	&        $z$   &$\theta_{m}$&$K$ 	&$K$ &$\delta K$ 	& Class \\
      &    	&     	       &  (\arcs)&$(\theta_{m})$ 	&$(63.9kpc)$	& &    \\ \hline
     1&       1&         1.1550&   1.5&    18.13 &     17.78&   30&  G \\
     2&       2&         0.9130&   1.0&    19.00 &     18.47&   19&  G \\
     3&       3&         0.7900&   2.5&    16.67 &     16.47&   15&  G \\
     4&       6&         1.0130&   2.5&    17.82 &     17.65&   27&  G \\
     5&       8&         1.5880&   1.5&    19.35 &     19.01&   34&  G \\
     6&       7&         0.5470&   2.5&    16.18 &     16.18&   16&  Q \\
     7&      10&         1.4370&   2.5&    17.79 &     17.79&   33&  Q \\
     9&      12&         0.2420&   2.5&    16.30 &     15.74&   25&  G \\
    10&      16&         1.0740&   2.5&    15.98 &     15.98&    7&  Q \\
    12&      18&         0.8210&   1.5&    19.13 &     18.74&   31&  G \\
    13&      20&         2.9500&   1.5&    19.86 &     19.49&   21&  G \\
    16&      23&         3.1260&   1.0&    19.85 &     19.32&   41&  G \\
    17&      24&         0.8930&   2.5&    18.02 &     17.84&   30&  G \\
    18&      25&         0.1090&   4.5&    13.02 &     12.45&   23&  G \\
    24&      32&         3.4310&   1.5&    19.69 &     19.30&   31&  G \\
    27&      44&         0.4230&   4.5&    15.87 &     15.78&   23&  G \\
    28&      38&         0.4720&   4.5&    15.98 &     15.91&   11&  G \\
    29&      39&         0.9650&   2.5&    17.29 &     17.29&   23&  Q \\
    32&      43&         1.1510&   1.5&    17.91 &     17.56&   23&  G \\
    33&      45&         1.2030&   2.5&    18.91 &     18.75&   27&  G \\
    35&      48&         0.4730&   4.5&    16.53 &     16.46&   20&  G \\
    37&      52&         0.5110&   1.5&    19.45 &     19.45&   26&  Q \\
    38&      53&         2.1160&   2.5&    17.25 &     17.25&   15&  Q \\
    39&      54&         1.5720&   2.5&    17.63 &     17.63&   33&  Q \\
    41&      58&         0.2950&   4.5&    15.07 &     14.89&   21&  G \\
    43&      64&         0.7780&   1.5&    17.54 &     17.15&   24&  G \\
    44&      62&         0.7900&   2.5&    17.98 &     17.98&   30&  Q \\
    45&      66&         0.7960&   2.5&    17.04 &     16.84&   13&  G \\
    46&      65&         0.7180&   2.5&    17.15 &     16.94&   13&  G \\
    47&      63&         0.5080&   4.5&    16.50 &     16.45&   21&  G \\
    49&      67&         0.4100&   4.5&    15.88 &     15.78&   25&  G \\
    52&      72&         1.6245&   1.0&    18.93 &     18.43&   22&  G \\
    53&      76&         0.4260&   2.5&    15.60 &     15.24&   22&  G \\
    54&      74&         0.4100&   4.5&    14.25 &     14.15&   11&  G \\
    55&      71&         0.5570&   4.5&    16.67 &     16.64&   15&  G \\
    62&      84&         0.5740&   1.5&    17.41 &     16.81&   28&  G \\
    63&      88&         0.3020&   4.5&    16.25 &     16.08&   25&  G \\
    65&      87&         0.5490&   2.5&    16.95 &     16.65&   26&  G \\
    67&      89&         0.4280&   4.5&    16.05 &     15.96&   18&  G \\
    68&      91&         0.5140&   1.5&    16.91 &     16.28&   22&  G \\
    72&      97&         2.4270&   2.5&    18.04 &     17.88&   21&  G \\	
    75&      98&         0.2650&   2.5&    15.32 &     14.80&   20&  G \\
    76&     102&         0.2820&   4.5&    15.28 &     15.09&   31&  G \\
    77&     104&         1.5120&   1.5&    19.01 &     18.67&   30&  G \\
    79&     106&         1.2550&   4.5&    17.37 &     17.44&   25&  G \\
    80&     110&         0.3660&   4.5&    14.77 &     14.65&   15&  G \\
    83&     116&         0.5210&   2.5&    16.48 &     16.17&   22&  G \\
    89&     117&         0.9090&   1.5&    19.53 &     19.16&   32&  G \\
    92&     122&         0.7430&   2.5&    16.47 &     16.47&   25&  Q \\
    93&     132&         0.1830&   1.5&    16.21 &     15.13&   31&  G \\
    95&     123&         0.0450&   4.5&    13.25 &     12.08&   27&  S \\
    96&     131&         2.7060&   1.0&    20.59 &     20.07&   27&  G \\
    99&     133&         0.7380&   4.5&    16.18 &     16.20&   21&  G \\
   102&     134&         0.4680&   2.5&    15.97 &     15.63&   28&  G \\
   105&     138&         3.3770&   1.0&    20.70 &     20.16&   36&  G \\
\hline								        
\end{tabular}
\caption{Aperture corrected $K$ band magnitudes for all sources with a measured spectroscopic redshift (presented in Paper 3) and a $K$--band detection.
Columns list the CENSORS and EIS identifiers; the redshifts; the aperture radius used to measure the $K$--band magnitude; the measured aperture magnitude; the magnitude after correction to 63.9kpc; the error on the corrected magnitude (in units of 0.01 mags) and the classification of the sources as radio galaxies (G), quasars (Q) or star forming galaxies(S). Paper 3 discusses the evidence upon which these classifications are made.
\label{Kcorrected_tab}}
\end{center}
\end{table}
\end{scriptsize}

\begin{scriptsize}
\begin{table}
\addtocounter{table}{-1}
\begin{center}
\begin{tabular}{|l|l|l|l|l|l|l|l|}
\hline
CEN & EIS & $z$ &$\theta_{m}$  &$K$ 		&$K$ 		&$\delta K$ 	& Class \\ 
    &    &     &  (\arcs)      &$(\theta_{m})$ 	&$(63.9kpc)$	&	&    \\ \hline
   107&     148&         0.5120&   2.5&     16.32&     16.01&   17&  G \\
   108&     153&         0.2300&   4.5&     14.93&     14.68&   23&  G \\
   111&     149&         0.4110&   4.5&     15.94&     15.84&   24&  G \\
   114&     166&         1.4260&   1.5&     19.23&     19.23&   34&  Q\\
   115&     155&         0.5450&   4.5&     15.22&     15.18&   23&  G\\
   116&     143&         2.6370&   2.5&     18.23&     18.23&   27&  Q\\
   117&     165&         1.2040&   2.5&     18.31&     18.15&   33&  G\\
   118&     161&         2.2940&   1.0&     19.82&     19.31&   22&  G\\
   120&     159&         2.8290&   1.5&     17.68&     17.31&   24&  G\\
   121&     164&         0.2460&   4.5&     14.86&     14.64&   20&  G\\
   122&     156&         0.2500&   4.5&     14.62&     14.40&   22&  G\\
   127&     186&         0.9220&   2.5&     17.28&     17.10&   24&  G\\
   129&     170&         2.4210&   1.0&     19.51&     19.00&   36&  G\\
   131&     169&         0.4700&   1.0&     16.84&     15.87&   33&  G\\
   134&     182&         2.3540&   1.5&     20.28&     19.93&   14&  G\\
   137&     187&         0.5260&   2.5&     16.84&     16.53&   26&  G\\
   138&     177&         0.5080&   2.5&     17.35&     17.03&   35&  G\\
   140&     199&         0.2650&   2.5&     16.19&     15.67&   33&  G\\
   141&     189&         2.8290&   1.0&     19.57&     19.05&   27&  G\\
   144&     179&         0.6960&   2.5&     17.50&     17.28&   32&  G\\
   145&     137&         0.4000&   4.5&     15.97&     15.87&   11&  G\\
   149&     185&         0.0290&   2.5&     13.95&     12.58&   35&  S\\
\hline
\end{tabular}
\caption{cont.}
\end{center}
\end{table}
\end{scriptsize}

Of the 142 CENSORS sources which are imaged in the $K$-band, 78 are detected and have spectroscopic redshifts (which are presented in Paper 3).
Using these redshifts, the observed $K$--band magnitude may be corrected to a consistent, physical radius thus allowing direct comparison of the magnitudes of the sources.

The $K$ band magnitudes presented here were measured in aperture of radius 1\arcsec, 1.5\arcsec, 2.5\arcsec and 4.5\arcsec.
Ideally a very large aperture would be used in order to be sensitive to the whole galaxy.
However the level of the sky background will affect the maximum radius aperture that is possible. 
In order to compare the magnitudes of galaxies, the magnitudes are corrected to an aperture of 63.9kpc (chosen for consistency with previous work).
Since most low redshift radio galaxies reside in large elliptical galaxies, the emission from the host galaxies will follow a curve of growth which can be used to correct the magnitudes.
The curve of growth presented in \cite{Sandage72} has been used to correct galaxies with redshift $z < 0.6$.
High redshift galaxies don't always have the same structures as those at low redshift so this method is not appropriate for them and so, following \cite{Eales97}, it is assumed that emission from higher redshift galaxies, measured within an aperture of radius, $r$, is proportional to $r^{\alpha}$, where $\alpha = 0.35$.

Where there was a choice of aperture, corrections were calculated for all. The corrected magnitude with the least error  (including an estimated 15\% error on the correction term, in order to account for an uncertainty in making the correction) was used.
This also allowed a check that aperture corrections agreed for different sized apertures, within the errors.
Aperture sizes below 2.5\arcs have only been used when either a contaminating galaxy is present, or the radio galaxy is so faint that larger apertures give photometric errors in excess of 0.35 magnitudes.
No corrections are applied for  quasar $K$--band magnitudes, as they are point-like sources.

The corrected K band magnitudes are presented in Table \ref{Kcorrected_tab}.
This table also lists the  quasars and star-forming galaxies that are identified in Paper 3.

\section{The CENSORS $K$--z Relation}
\label{kz}

\begin{figure}

\centerline{\psfig{file=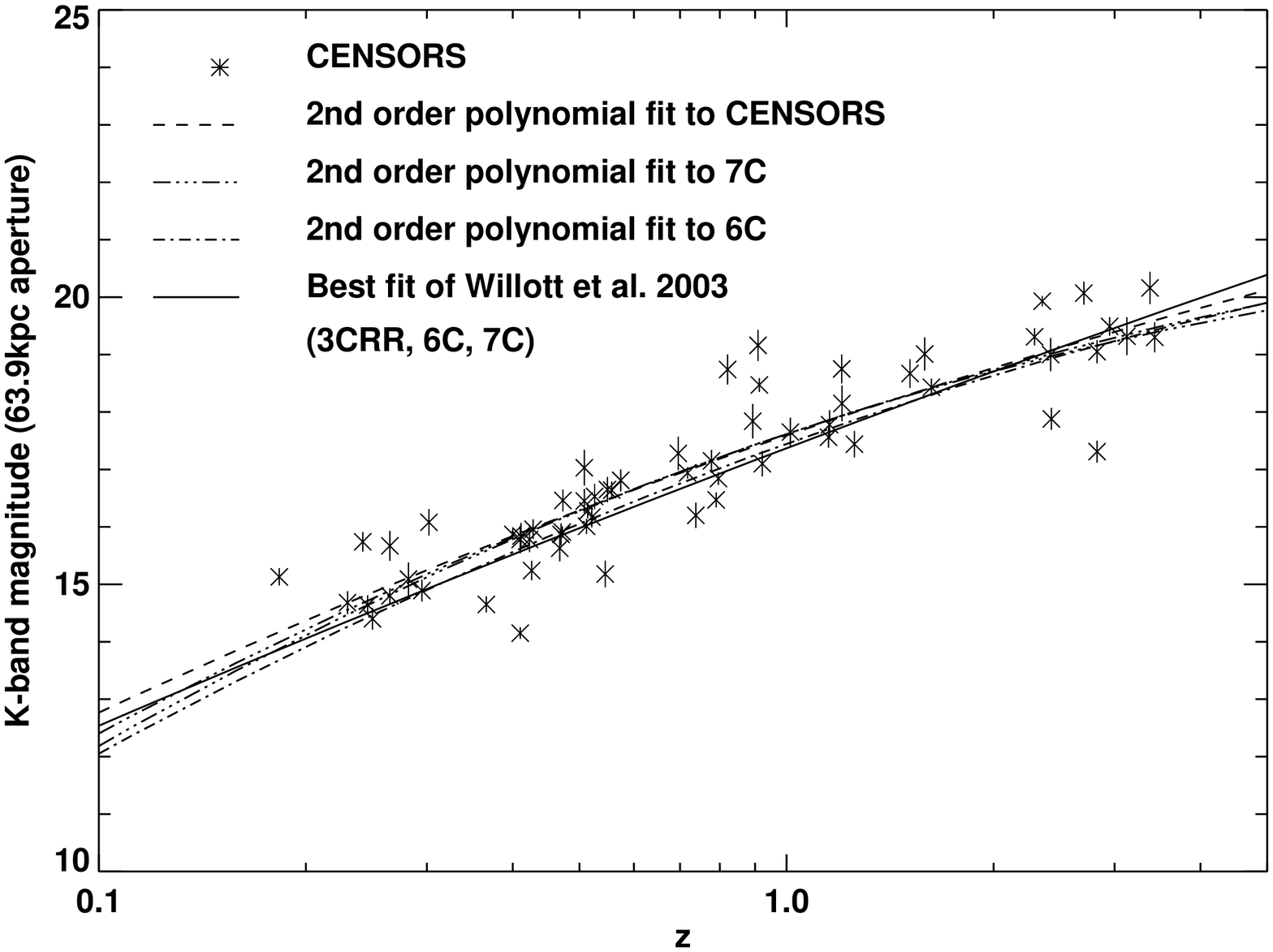,angle=90,width=\linewidth}}
\caption[The $K$--$z$ relation for radio galaxies]{$K$-band magnitude against redshift for all CENSORS source for which both are available, excluding star forming galaxies and quasars.
The solid line shows the best fitting relation of \cite{Willott03KZ} which is a 2nd order polynomial in  $\log z$, derived using the 3CRR, 6CE, 6C* and  7C regions I, II and III.
The dot-dash line shows the $K$--$z$ relation as derived for the 6C samples used by \cite{Willott03KZ} and the dash-triple-dot line shows that of the 7C samples used by \cite{Willott03KZ}.
The dashed line is the best fitting relation based upon the CENSORS sample alone.
The 6C, 7C and CENSORS fits are also polynomials of 2nd order in $\log z$.
\label{CENSORSKZ_ps}}
\end{figure}

As described in the Introduction, the $K$--$z$ relation is known to vary for radio selected samples of different flux-density limit.
Previous work with the 3CRR, 6C and 7C surveys has shown that whilst the 6C and 3CRR samples share a similar $K$--$z$ relation at low redshifts, the 6C sample, which is \squig~~5 times fainter than 3CRR, is 0.6 magnitudes fainter at $z > 1$ (see, for example, \citeauthor{LL84} 1984, \citeauthor{EalRaw96} 1996 and \citeauthor{Inskip02} 2002).
\cite{Willott03KZ} showed that the 7C sample (which is a further four times fainter in radio flux-density limit than 6C) is fainter in the $K$--band than 3CRR at all redshifts and speculates that the 6C sample, which is sparse in objects at the lowest redshifts, and therefore carries a large error, may be consistent with being fainter in $K$ at low redshifts too.

The spectroscopic completeness of the CENSORS sample is  60\%, and so it is possible to make a comparison of the CENSORS $K$--$z$ relation against this previous work. 
This is done whilst noting that strong conclusions should be delayed until the spectroscopic completeness is higher.

Figure \ref{CENSORSKZ_ps} plots the aperture corrected $K$-band magnitude against redshift for all sources for which both these variables have been measured; excluding two star forming galaxies (CENSORS 95 and 149) and 11 quasars (CENSORS 6, 7, 10, 29, 37, 38, 39, 44, 92, 114 and 116) as these will be hosted by different galaxies. 
The identification of the quasars and star forming galaxies is largely based upon spectroscopic evidence and is discussed in full detail in Paper 3. 
Overplotted are the $K$--$z$ relations from: \cite{Willott03KZ} (solid line, based upon 3CRR, 6C and 7C); the $K$--$z$ relation based upon the 6CE and 6C* samples alone (dash-dot); the 7C sample (regions I,II and III) alone (dash-triple-dot) and the CENSORS points (dashed).
All fits are 2nd order polynomials in $\log z$ and the 6C and 7C samples used are identical to those used in \cite{Willott03KZ}.

The Willott fit is clearly a reasonable fit to the CENSORS, however the CENSORS best fit is fainter at low redshifts.
The 7C fit is given by:
\begin{equation}
K = 17.62 + 4.04 \log z - 1.39 \log^{2} z,
\end{equation}
and shows that in fact the CENSORS sample is in better agreement with the 7C sample rather than the 3CRR. Note that no fit to the CENSORS data points is given due to the incomplete spectroscopic information for this sample at present.
\begin{figure}
\begin{tabular}{c}
\centerline{\psfig{file=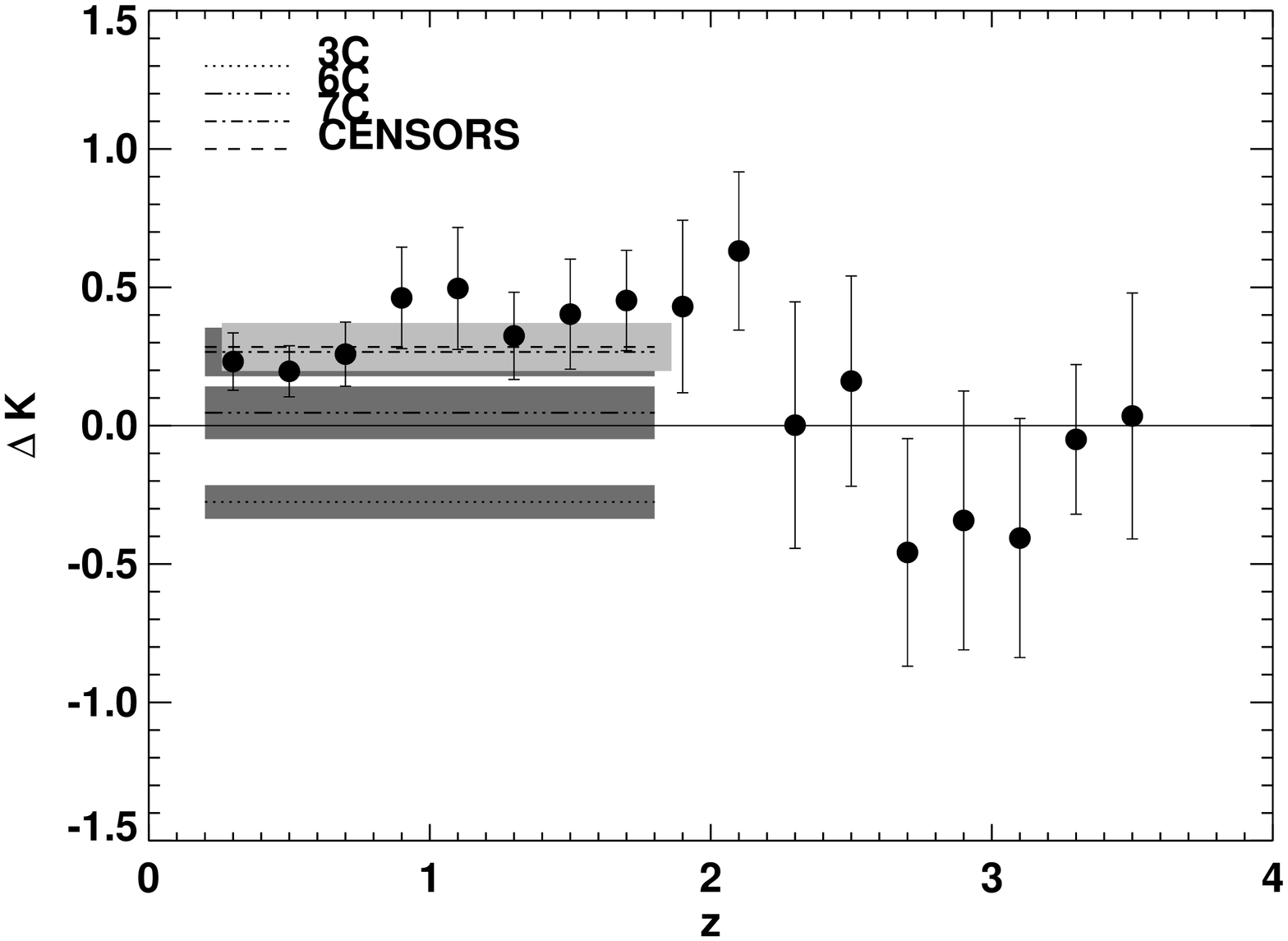,angle=90,width=\linewidth}}\\
{\bf (a)}\\
\centerline{\psfig{file=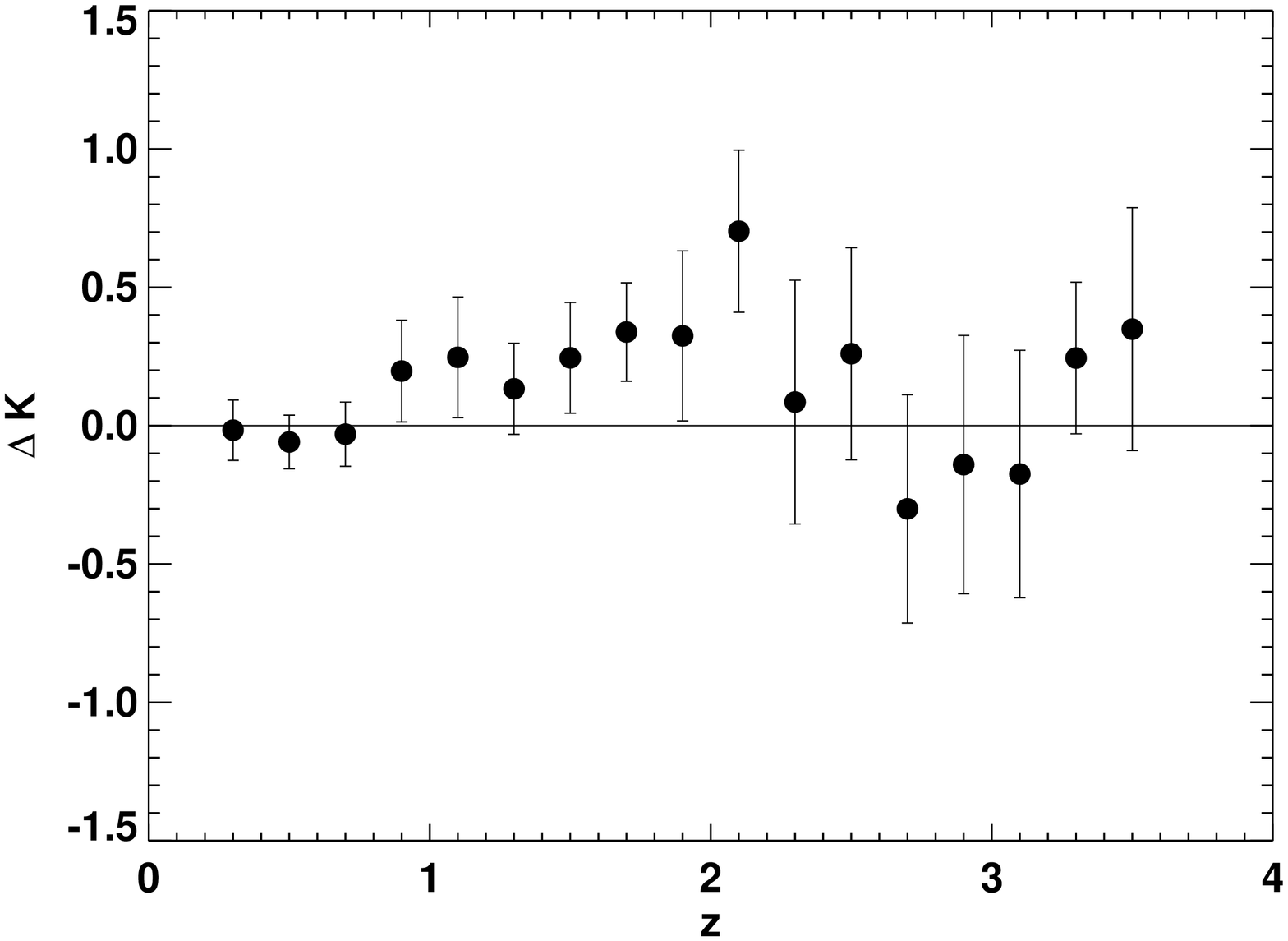,angle=90,width=\linewidth}}\\
{\bf (b)}\\
\end{tabular}
\caption[]{
{\bf (a)} The  offset, $\Delta K$, of the CENSORS sources (where both a $K$--band magnitude and spectroscopic redshift are available, and excluding star-forming sources and quasars) compared to the best fitting  $K$--$z$ relation for radio galaxies from \cite{Willott03KZ} (based upon 3CRR, 6CE, 6C* and 7C, I,II and II).
Here $\Delta K = K - K_{fit}$.
Overplotted are the mean $\Delta K$ values for the 3CRR (dotted), 6C (dash-triple-dot), 7C(dash-dot) and CENSORS (dashed) in the range $0.2 < z < 1.8$. 
The shaded regions indicate the 1-$\sigma$ standard errors about the means and the shaded region for the CENSORS sample has been offset 0.06 in $z$ in order that the extent of the 7C standard errors may be displayed.
This plot is based upon Figure 3 of \cite{Willott03KZ} and the relevant data were provided by Willott (private communication).
The data are investigated in  bins of width $z = 0.62$ plotted  in steps of $\Delta z = 0.2$.
The errors bars show the standard errors and are correlated as a result of the overlapping bins.

{\bf (b)} Shows the $K$--band offsets of the CENSORS sources as compared to the $K$--$z$ relation based upon 7C only.

\label{CENSORSoffsets_ps}

}
\end{figure}
These low redshift differences are investigated in Figure \ref{CENSORSoffsets_ps} {\bf (a)} which plots the difference between the CENSORS sample and the best fitting $K$--$z$ relation of \cite{Willott03KZ}.
This Figure replicates Figure 3 of \cite{Willott03KZ} but uses bins of width $z = 0.62$, rather than 0.6, plotted in steps of $\Delta z = 0.2$.
The change in bin size ensures that every bin has more than one source.
The errors bars show the standard errors and are correlated as a result of the overlapping bins.
This figure shows that CENSORS is fainter than 3CRR over all redshifts and by a similar magnitude as the 7C sample.
At redshifts greater than \squig~~2 there appears to be little difference between the CENSORS points and the best fit. 
However at these redshifts there are no 3CRR sources and so the best fit is essentially a fit to the 6C and 7C samples alone.
This is further demonstrated in Figure \ref{CENSORSoffsets_ps} {\bf (b)} which plots the $K$--band offsets of the CENSORS sources from the 7C $K$--$z$ relation.

This indicates that it is the 3CRR $K$--$z$ relation that should be considered `anomalous' as the other surveys tend to agree with a more consistent relation.
The simplest interpretation of this would be that the most powerful radio sources, as sampled by the 3CRR sample, have additional $K$--band light either directly or indirectly associated with te AGN activity (e.g. through associated star formation).
However, at lowest redshifts, the 3CRR sources are not believed to be significantly AGN contaminated (e.g. \citeauthor{BLR98} 1998).
So it may be wiser to consider these data as a constraint on the link between $K$--band magnitude and radio luminosity, via galaxy mass and black hole mass, given variations in the  accretion rate.
These conclusions however are made with only 60\% spectroscopic completeness in the CENSORS sample and so a full analysis must be left until that figure has increased.
At $z < 1$, redshifts are obtained from emission lines or features in bright continua, therefore, the spectroscopically identified sources may be biased to be brighter objects. 
However at these redshifts our spectroscopic completeness is believed to be quite high ($\ge 80$\%, see Paper 3).
Higher redshift objects tend to be spectroscopically identified by emission lines alone and so may be less biased to bright galaxies, though they will still suffer bias due to secondary effects.

Despite these difficulties, given that  our spectroscopic completeness is likely to be higher at low redshifts than at high redshifts, the result that CENSORS is fainter in the $K$--band than 3CRR and of similar magnitude to 7C seems likely to stand upon further investigation.
In the meantime it is sufficient to conclude that the CENSORS $K$--$z$ relation is not substantially different from the best fit to the 7C data and that relation may used to estimate redshifts for the non-spectroscopically identified CENSORS host galaxies.

\section{Summary}
\label{summary}

Of the 150 CENSORS sources, 142 have been observed in the $K$--band  using the AAT, UKIRT and the  VLT over the last three years.
These images have been used to identify the host galaxies of CENSORS sources which lack an identification in the EIS $I$--band images and to provide $K$--band photometry for all CENSORS sources, where possible.

Of the eight sources not observed in the $K$--band, seven have $I$--band identifications of the radio source host galaxy and the remaining source has a weak $I$--band detection, which, when targeted spectroscopically, yielded a single emission line (see Paper 3)

There are eighteen sources for which a host galaxy is  not clearly detected.
However, of these, six are detected in the $I$--band. Of the remaining twelve some have complications, such as possible association with other radio sources or numerous outstanding candidate host galaxies, leading to uncertainty.
Low frequency radio data, obtained at 610 and 325MHz, may help to resolve issues where interpretation of the radio data is difficult.
Four sources remain  undetected  are likely to be cases of high redshift host galaxies, and  deeper infrared images are required.

Aperture corrections have been calculated for those sources with a $K$--band magnitude presented in this paper and a spectroscopic redshift presented in Paper 3.
From these a $K$--$z$ relationship for the radio galaxies of the CENSORS sample has been established and shown to be
offset to fainter $K$--band magnitudes over the range $0.2 < z < 1.8$ with respect to the best fitting $K$--$z$ relation of \cite{Willott03KZ}.
Since the $K$--band magnitudes of 3CRR sources (at low redshift) are not believed to be significantly affected by the AGN, this provides constraints on any model which attempts too explain the $K$--$z$ relation in terms of black hole mass and accretion rate.
However these conclusions should be taken on the understanding that a higher spectroscopic completeness, in the CENSORS sample, is required before it is considered certain.

\section*{Acknowledgements} 

MHB is grateful for the support of a  PPARC research studentship.
Observations were made using the United Kingdom Infrared Telescope which is operated by the Joint Astronomy Centre on behalf of the U.K. Particle Physics and Astronomy Research Council.
Further observations were carried out using  the Anglo-Australian Telescope and the ESO Very Large Telescope at the Paranal observatory (Program-ID number 69.A-0047).

\label{lastpage}
\bibliography{CENSORS_Kband.bib} 
\bibliographystyle{mn2e} 
\clearpage
\onecolumn

\appendix
\section{$K$--band images of CENSORS sources}

\begin{figure}
%\vspace*{-14.5cm}
%\centerline{\psfig{file=images1.ps,angle=0,width=15cm,clip=}}
\caption{{\it See the attached jpg files.} CENSORS $K$-band images are shown in greyscale. Some images have been Gaussian smoothed for clarity. 
The contours plot the radio maps (presented in Paper 1) based upon VLA observations in BnA and CnB array at 1.4GHz. The resolution of the radio maps is 7 arcsec or better, and in many cases reaches 3 or 4 arcsec (see Paper 1 for details).\label{kband_ps}}
\end{figure}

\begin{figure}
%\begin{tabular}{cc}
%\hspace*{2.7cm}I$-Band             & \hspace*{2.7cm}$K$-band   \\
%\psfig{file=eisd16_comblab_bb.ps,angle=0,width=7cm,height=10.5cm,clip=}&
%\psfig{file=cen103+10+90_label.eps,angle=0,width=7cm,height=10.cm,clip=}\\
%\end{tabular}
%\bigskip
%\bigskip
%\bigskip
%\begin{tabularx}{\linewidth}{XXX}
%\multicolumn{3}{c}{$K$-Band}\\
%\psfig{file=cen103+10+90_k_ov_C2.eps,angle=0,width=3cm,clip=}&
%\psfig{file=cen103+10+90_k_ov_D2.eps,angle=0,width=3cm,clip=}&
%\psfig{file=cen103+10+90_k_ov_F2.eps,angle=0,width=3cm,clip=}\\
%\hspace*{1.6cm}C & \hspace*{1.6cm}D & \hspace*{1.6cm}F \\
%\multicolumn{3}{c}{$I$-Band}\\
%\psfig{file=eisd16b_c_over.ps,angle=0,width=3cm,clip=}&
%\psfig{file=eisd16b_d_over.ps,angle=0,width=3cm,clip=}&
%\psfig{file=eisd16b_f_over.ps,angle=0,width=3cm,clip=}\\
%\hspace*{1.6cm}C & \hspace*{1.6cm}D & \hspace*{1.6cm}F \\
%\end{tabularx}
\caption{{\it See attached jpg files.} $I$--band \citep{CENSORS1} and $K$--band images (this work) of CENSORS 10, 90 and 103.\label{censors103_fig}}
\end{figure}

\end{document}